\documentclass[aps,preprint]{revtex4}
\usepackage[dvips]{graphicx}
\usepackage{amsmath}
\usepackage{amssymb}
\usepackage{graphicx}
\usepackage{mathrsfs}
\usepackage{ulem}
\usepackage{times}
\makeatletter
\def\bib@device#1#2{}
    \newcommand{\rmnum}[1]{\romannumeral #1}
    \newcommand{\Rmnum}[1]{\expandafter\@slowromancap\romannumeral #1@}
  \makeatother

\begin{document}
\title{A Local Mott Transition in NiO Resistance Random Access Memory}
\author{Kan-Hao Xue}
\email{xuekanhao@gmail.com}
\affiliation{Department of Electrical and Computer Engineering, University of Colorado, Colorado Springs, Colorado 80918, USA}
\affiliation{Symetrix Corporation, 5055 Mark Dabling Boulevard, Colorado Springs, Colorado 80918, USA}
\author{Carlos A. Paz de Araujo}
\affiliation{Department of Electrical and Computer Engineering, University of Colorado, Colorado Springs, Colorado 80918, USA}
\affiliation{Symetrix Corporation, 5055 Mark Dabling Boulevard, Colorado Springs, Colorado 80918, USA}
\author{Jolanta Celinska}
\affiliation{Symetrix Corporation, 5055 Mark Dabling Boulevard, Colorado Springs, Colorado 80918, USA}
\author{Christopher McWilliams}
\affiliation{Symetrix Corporation, 5055 Mark Dabling Boulevard, Colorado Springs, Colorado 80918, USA}
\begin{abstract}
The physics of the Mott transition in the anodic region of NiO Resistance Random Access Memory (RRAM) is discussed from the Hubbard model. The Hubbard approximation is examined in details and it is shown that the Wannier functions in the definition of Hubbard $U$ should not be replaced by atomic $s$-functions when it comes to the metallic side of the transition. The corresponding effective Hubbard $U$ is subject to variations, which may also be understood by introducing an effective permittivity of the solid as an $ansatz$. Furthermore, the transition could be demonstrated in the Brinkman-Rice picture. Finally, the anodic characteristics of such transition show that it is a local Mott transition. Therefore, the unipolar switching NiO RRAM can still have asymmetric I---V curves when a capping layer is inserted, which is explained qualitatively by quantum mechanics.   
\end{abstract}
\maketitle
The Mott metal-insulator transition (MIT) involves collective localization or de-localization of electrons in a solid \cite{1,2}. Practically, Mott transition is usually achieved by adding pressure or doping some ions of different sizes. Nevertheless, none of these two approaches could be dynamical in an electronic device such that Mott transition can serve as the device physics of a resistive switching memory in consumer electronics. On the other hand, there has been an electronic device Resistance Random Access Memory (RRAM) that utilizes transition metal oxides (TMO) such as NiO \cite{3}, which can be subject to a Mott transition theoretically. Rozenberg $et\ al.$ proposed that the boundary domains of a NiO RRAM could suffer from a Mott MIT because of strong correlation effects \cite{4}-\cite{6}. They further pointed out that the RESET transition is due to a lowered occupation in the bottom domain \cite{6}. However, the mechanism of such a transition was not discussed in details, and they predicted that the insulating region in a unipolar switching NiO RRAM should be near the cathode (bottom domain) \cite{6}, which is against several experimental findings \cite{7,8}.

In a recent paper \cite{9} we established the analytical current-voltage formulae for carbonyl doped NiO RRAM by assuming the densities of states in NiO are different on the metal and insulator sides. Nevertheless, the RESET transition is beyond the scope of those quantum transport equations. Even if SET seems a phenomenon like tunneling or breakdown, the RESET phenomenon is more mysterious since it is an increased applied voltage that causes a transition from ON state to OFF state. This is different from negative differential resistance (NDR) phenomenon because no orders of resistance change can be observed in NDR. According to this peculiarity, many groups have attributed RESET to the thermal rupture of conduction filaments \cite{10}-\cite{13}, because an increased voltage leads to stronger Joule heating. Nevertheless, such explanation has difficulty in that the power through the device can be higher at the SET point than the RESET point and one cannot explain why the filaments are not ruptured right after SET \cite{9}. Our starting point of the RRAM model was the fact that the insulating region in NiO RRAM only lies near the anode, as confirmed by the following experiments: (\rmnum{1}) the sputtering damage experiment carried out by Kinoshita $et\ al.$ \cite{7}; (\rmnum{2}) the transmission electron spectroscopy (TEM) observation by Park $et\ al.$ \cite{8}. These two experiments, especially the first one, strongly confirm that the local region near the anode must be emphasized if the RESET mechanism is to be clarified. Note that near the anode, the electron quasi-Fermi level is lowered during the transport and thus there should be a dynamical electron deficit occurring near the anode. This implies that the Mott criterion \cite{14} could play a role, which states that a solid is an insulator if the free electron concentration ($n_f$) is so low that
\begin{equation}
{n_f}^{1/3}a_B < 0.26\ \ .
\end{equation}
Here $a_B = 4\pi\epsilon\hbar^2/me^2$ is the effective Bohr radius. When it comes to RESET, the applied voltage causes strong Fermi level splitting in the two electrodes and the suppression of electron concentration near the anode may incur a local Mott transition. This conjecture is attractive in that the Mott transition can be found and utilized in real electronic devices. Yet, notwithstanding its successful explanation to RESET, the mechanism of such Mott transition deserves further discussions. In the present paper, this local Mott transition will be examined from the well-known Hubbard model and the Brinkman-Rice picture.

The Hubbard model is the general model in discussing narrow $3d$ band TMOs as well as the Mott transition. In writing down the Hamiltonian of a solid, it neglects electron-phonon interactions and the ionic Hamiltonian. Furthermore, the Wannier representation is preferred such that the Hamiltonian writes as:
\begin{equation}
H = \sum\limits_{ij\sigma}T_{ij}a_{i\sigma}^\dagger{a}_{j\sigma} + \frac{1}{2}\sum\limits_{ijkl}^{\sigma,\sigma'}v\left(ij;kl\right)a^\dagger_{i\sigma}a^\dagger_{j\sigma'}a_{l\sigma'}a_{k\sigma}\ \ ,
\end{equation}
where
\begin{equation}
T_{ij} = \int{d\vec{r}}\ \omega^*\left(\vec{r} - \vec{R}_i\right)\mathfrak{h}(\vec{r})\ \omega\left(\vec{r} - \vec{R}_j\right) = \frac{1}{N}\sum\limits_{\vec{k}}\mathcal{E}\left(\vec{k}\right)e^{\mathrm{i}\vec{k}\cdot\left(\vec{R}_i - \vec{R}_j\right)}
\end{equation}
is the hopping integral, $\omega$ represents Wannier functions, and $\mathfrak{h}$ is the single electron Hamiltonian; $v\left(ij;kl\right)$ can be expressed in terms of Wannier functions as:
\begin{equation}
v\left(ij;kl\right) = \frac{e^2}{4\pi\epsilon_0}\iint{d}\vec{r}_1{d}\vec{r}_2\frac{\omega^*\left(\vec{r}_1 - \vec{R}_i\right)\omega^*\left(\vec{r}_2 - \vec{R}_j\right)\omega\left(\vec{r}_2 - \vec{R}_l\right)\omega\left(\vec{r}_1 - \vec{R}_k\right)}{\left|\vec{r}_1 - \vec{r}_2\right|}\ \ .
\end{equation}
There are many types of integrals in $v\left(ij;kl\right)$, which renders the Hamiltonian unsolvable. Hubbard in his original work noted that the Wannier functions could be approximately replaced by atomic $s$-functions ($\phi$) when it comes to a very narrow band \cite{15}. Hence, he assumed that only the single-center integral $v\left(ii;ii\right)$ is significant and should be retained. It has another name as the Hubbard $U$:
\begin{equation}
U = v\left(ii;ii\right) = \frac{e^2}{4\pi\epsilon_0}\iint{d}\vec{r}_1{d}\vec{r}_2\frac{\phi^*\left(\vec{r}_1 - \vec{R}_i\right)\phi^*\left(\vec{r}_2 - \vec{R}_j\right)\phi\left(\vec{r}_2 - \vec{R}_l\right)\phi\left(\vec{r}_1 - \vec{R}_k\right)}{\left|\vec{r}_1 - \vec{r}_2\right|}\ \ .
\end{equation}
By neglecting multi-center integrals, one arrives at the Hubbard model. Later, Brinkman and Rice \cite{16} started from this model and utilized a variational method by Gutzwiller \cite{17}, showing that a second order MIT could be predicted. Their scenario is that the Hubbard $U$ (force of localization) and the tight binding bandwidth $W$ (force of delocalization) are competing with each other. A solid with half-filled $3d$ valence band, such as NiO, can be an insulator if the Hubbard $U$ overwhelms the bandwidth $W$. According to the definition of $U$ [Equation (5)], it should be treated as a constant. Therefore, the only way to achieve an insulator-to-metal transition in NiO is to add pressure such that $W$ would increase. It then implies that a Mott transition is not practical in NiO RRAM.

Nevertheless, the above discussions are all based on the Hubbard approximation. As pointed out by Herring \cite{18}, Friedel \cite{19}, and Cyrot \cite{20}, the practical Hubbard $U$ should not be the atomic integral, but rather an effective one. We here argue that the Hubbard approximation can be divided into two steps: (\rmnum{1}) to neglect all the multi-center integrals; (\rmnum{2}) to replace the Wannier functions in $v\left(ii;ii\right)$ by atomic functions. In writing Equation (5), one applies the whole two-step approximation. However, the argument that $U$ is a constant strongly depends on step (\rmnum{2}). If only the first step is applied, then
\begin{equation}
U = \frac{e^2}{4\pi\epsilon_0}\iint{d}\vec{r}_1{d}\vec{r}_2\frac{\omega^*\left(\vec{r}_1 - \vec{R}_i\right)\omega^*\left(\vec{r}_2 - \vec{R}_j\right)\omega\left(\vec{r}_2 - \vec{R}_l\right)\phi\left(\vec{r}_1 - \vec{R}_k\right)}{\left|\vec{r}_1 - \vec{r}_2\right|}\ \ ,
\end{equation}
whose value is related to the Wannier functions. In case the corresponding energy band is not that narrow, $e.g.$ in a metal \footnote{It was shown in our previous works \cite{21} that extrinsic carbonyl ligand doping may lead to a metallic NiO in the virgin state. The fact that NiO becomes metallic implies that its $3d$ band is no longer very narrow and the tight binding approximation is subject to failure.}, the Wannier functions may have nothing to do with the atomic functions. If NiO is subject to an MIT, then the forms of Wannier functions are supposed to be very different on the two sides. Consequently, we prefer one-step Hubbard approximation for MIT. The difficulty lies in that the Wannier functions are unknown, otherwise through
\begin{equation}
\psi_{n\vec{k}}\left(\vec{r}\right) = \frac{1}{\sqrt{N}}\sum\limits_{j=1}^Ne^{\mathrm{i}\vec{k}\cdot\vec{R}_j}\omega_n\left(\vec{r} - \vec{R}_j\right)
\end{equation}
one can derive exactly the Bloch functions ($\psi$). In order to be consistent with the Brinkman-Rice picture, the following $ansatz$ can be proposed: assume there is an effective permittivity $\epsilon$ such that
\begin{equation}
U = \frac{e^2}{4\pi\epsilon}\iint{d}\vec{r}_1{d}\vec{r}_2\frac{\phi^*\left(\vec{r}_1 - \vec{R}_i\right)\phi^*\left(\vec{r}_2 - \vec{R}_j\right)\phi\left(\vec{r}_2 - \vec{R}_l\right)\phi\left(\vec{r}_1 - \vec{R}_k\right)}{\left|\vec{r}_1 - \vec{r}_2\right|}\ \ .
\end{equation}
\begin{figure}[!b]
  \centering
  \includegraphics[bb=0 0 444 514,width=3.5in,height=4.05in,keepaspectratio]{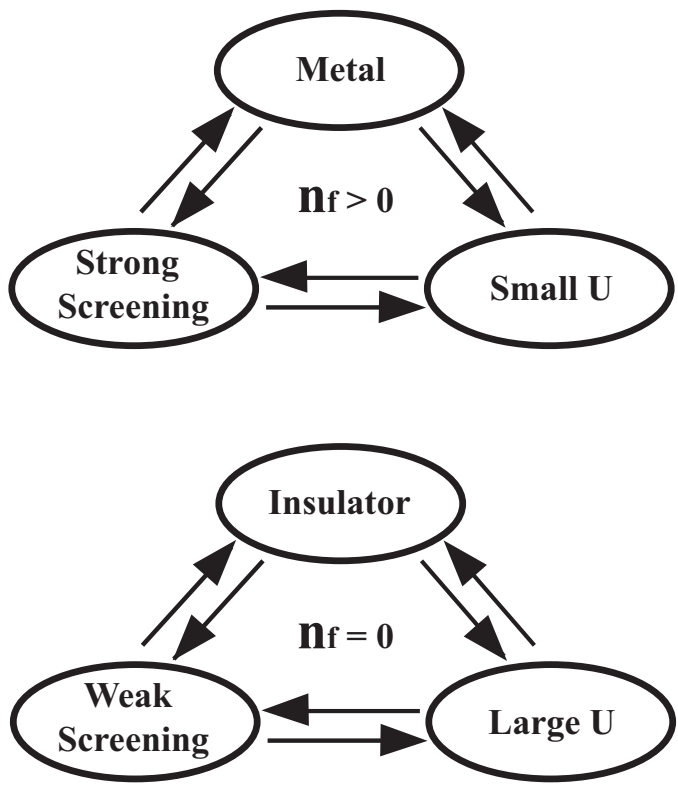}
  \caption{The interrelated criteria for a metal and an insulator in a NiO-like TMO, where $n_f$ is the free electron concentration.}
  \label{fig:Bond}
\end{figure}

The permittivity stands for screening effects related to long-range Coulomb interactions, which are totally neglected in the original Hubbard model. Such was previously pointed out by Cyrot \cite{20}, and Tsuda $et\ al$ \cite{22}. On the other hand, using this $ansatz$ it is natural that $U_{metal} < U_{insulator}$ because the charge screening is strong when some of the $3d$ electrons become itinerant. Also, one can envisage a first order MIT in NiO. The key parameter is the free electron concentration that determines the strength of charge screening. A larger $n_f$ leads to a larger effective permittivity, a smaller Hubbard $U$, and finally a higher conductivity. To sum up, the relations shown in Fig. 1 are followed.
\begin{figure}[!b]
  \centering
  \includegraphics[bb=0 0 515 264,width=5.67in,height=2.9in,keepaspectratio]{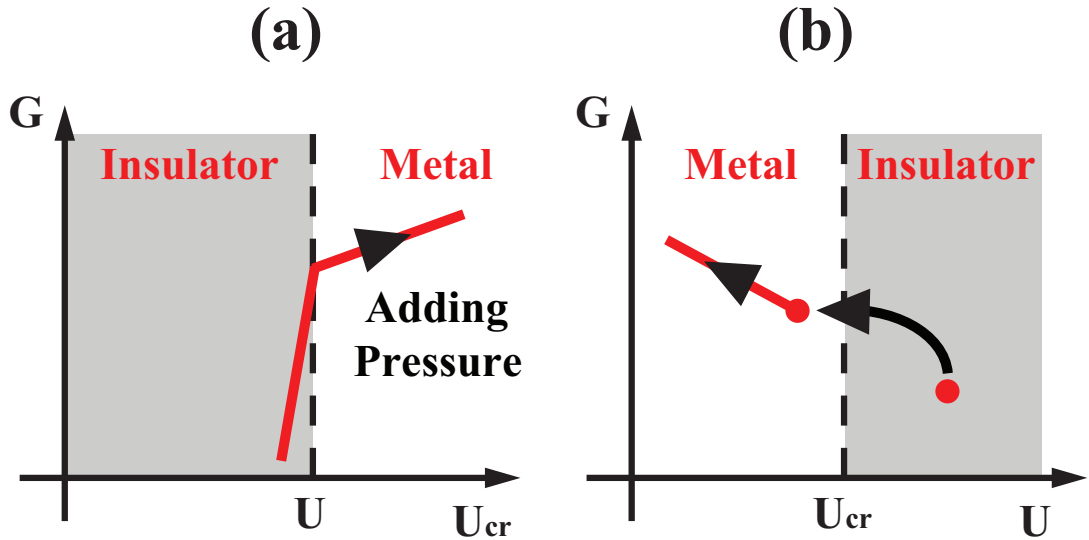}
  \caption{Illustration of the two ways to transform NiO into a metal: (a) by increasing $W$ and thus the critical value $U_{cr}$; (b) by decreasing the Hubbard $U$. Here $G$ represents the conductance.}
  \label{fig:TwoWays}
\end{figure}
\begin{figure}[!b]
  \centering
  \includegraphics[bb=0 0 524 454,width=4in,keepaspectratio]{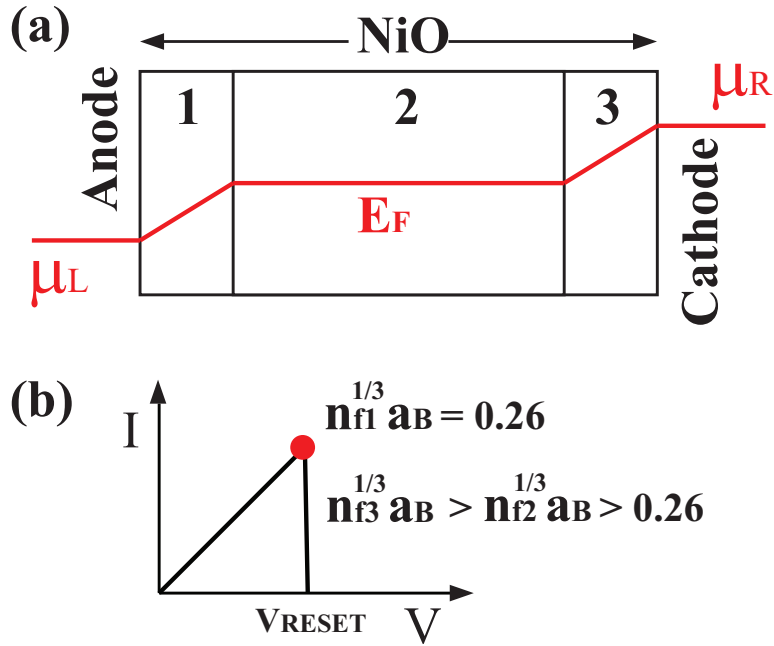}
  \caption{(a) The three regions in NiO RRAM and the average quasi-Fermi level through the device when a voltage is applied; (b) demonstration of the ON-state I---V and the RESET phenomenon.}
  \label{fig:Regions}
\end{figure}

Our argument is similar to that of Mott's 1949 paper \cite{23}. Mott claimed that once a few electrons get delocalized, there is no reason why no more electrons become delocalized because the existing itinerant electrons would function. He deduced that the transition should be of first order. On the other hand, the present paper claims that no intermediate states exist between the two sides of NiO, though NiO could be more metallic due to a further accumulation of electrons. The introduced $ansatz$ allows a schematic demonstration of such Mott transition within the Brinkman-Rice picture, as in Fig. 2. A traditional way to achieve Mott transition in NiO is by adding pressure, which is to increase $W$ and hence the critical value $U_{cr}$, while keeping $U$ constant. The present way, on the contrary, is to reduce $U$ while keeping $U_{cr}$ constant. 

Granted that the metal and insulator sides of NiO could be described from screening effects, another question arises how one can switch from one side to the other. There are two approaches to reduce the free electron concentration in NiO: (\rmnum{1}) to localize the itinerant $3d$ electrons; (\rmnum{2}) to reduce the total electron concentration. Approach (\rmnum{1}) faces a cyclic difficulty that once the $3d$ electrons get localized one already has a Mott transition, which requires a reduced $n_f$. Approach (\rmnum{2}) seems against Kirchhoff's law in the circuits. Nevertheless, in a non-equilibrium problem, there can be local electron deficit and excess at the boundaries of the device [shown in Fig. 3(a), assuming ballistic transport] \cite{9}. The total electron concentration near the anode is diminished as a voltage is applied, since the electron quasi-Fermi level there is lowered. The peculiar behavior of RESET, as illustrated in Fig. 3(b), is then explained as a sufficient separation of quasi-Fermi levels in the two electrodes. An interesting point here is the cathodic region. Due to the conservation of charges, that region must accommodate extra electrons without incurring further changes or phase transitions, such that the anodic region could suffer from the same amount of electron deficit. A higher free electron concentration tends to make the cathodic region more metallic. As the three regions of the device are in series, the resistance change near the cathode has almost no influence on the entire device's behavior.
\begin{figure}[!b]
  \centering
  \includegraphics[bb=0 0 307 268,width=3in,keepaspectratio]{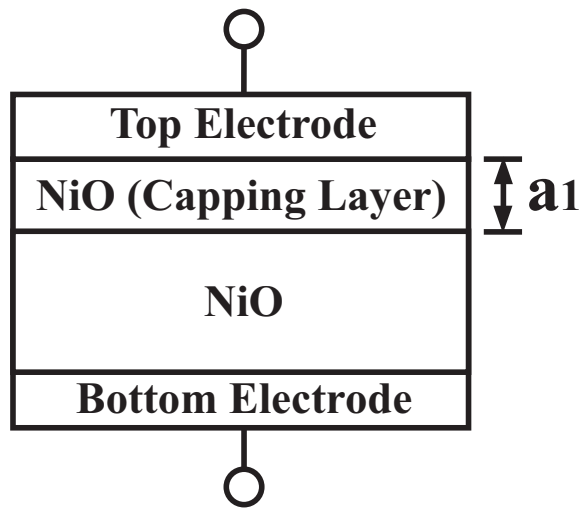}
  \caption{The device structure of a NiO RRAM with a single capping layer, where $a_1$ is the thickness of the capping layer.}
  \label{fig:Cap}
\end{figure}
\begin{figure}[!b]
  \centering
  \includegraphics[bb=0 0 285 462,width=3.5in,keepaspectratio]{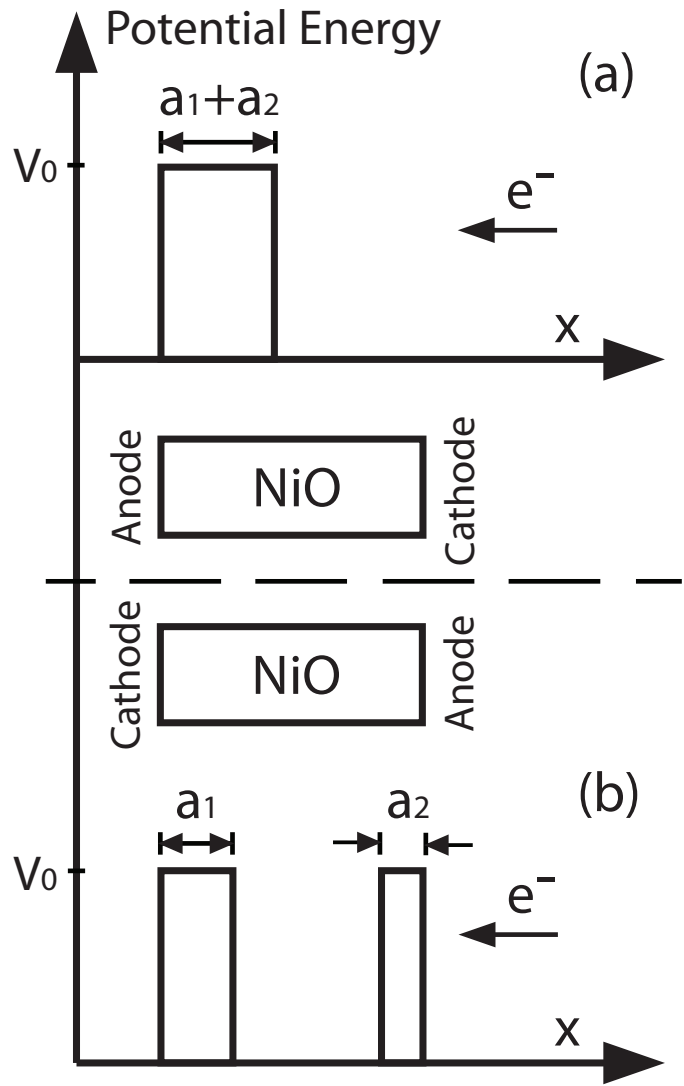}
  \caption{Schematic diagrams of the transport through a NiO RRAM with a capping layer: (a) positive polarity; (b) negative polarity.}
  \label{fig:Tunnel}
\end{figure}

That the Mott transition is called a local one should be understood within the framework of an electronic device. Admittedly, Mott transition is always due to collective electron behaviors rather than defective behaviors. Nevertheless, the possibility of such a Mott transition in an electronic device strongly depends on the local property of the anodic region. One of its great consequences is the asymmetric behavior of the device. The so-called ``unipolar" switching RRAM necessarily implies a symmetric I---V behavior. However, this symmetry can be broken once a voltage with certain polarity has been applied to RESET the device. A good example of this is to insert an insulating capping layer (undoped or less doped NiO) between carbonyl doped NiO and one particular electrode \cite{24}, as shown in Fig. 4. In the OFF state, the 1-D potential through such a structure depends on the polarity of the last RESET voltage. Figure 5 illustrates the potential profiles of: (a) positive polarity and (b) negative polarity. The interesting question is regarding the SET voltage, since this is a conventional tunneling problem in quantum mechanics. The transmission coefficient $T$ through a 1-D finite potential barrier $V_0$ is \cite{25}:
\begin{equation}
T(\mathcal{E}) = \left[1+\left(\frac{k^2 + \gamma^2}{2k\gamma}\right)^2sinh^2(2\gamma{a})\right]^{-1}\ \ ,
\end{equation}
where $k$ is the wave vector, $a$ is the thickness of the barrier, and
\begin{equation}
\gamma = \frac{\sqrt{2m\left(V_0 - \mathcal{E}\right)}}{\hbar}
\end{equation}
actually involves the energy of the electron. When the transmission coefficient is high, as is when SET occurs, the formula can be simplified as
\begin{equation}
T(\mathcal{E}) \simeq \frac{1}{1+k^2a^2}\ \ .
\end{equation}
In the cases (a) and (b), we may first assume the same energy of the incident electrons. The potential barriers of the capping layer and the insulating NiO generated during RESET are assumed to be the same:
\begin{equation}
T_a = \frac{1}{1+k^2\left(a_1 + a_2\right)^2}\ \ ,
\end{equation}
\begin{equation}
T_b = \frac{1}{1+k^2a_1^2}\frac{1}{1+k^2a_2^2}\ \ ,
\end{equation}
thus
\begin{equation}
\frac{T_a}{T_b} = 1 + \frac{k^2a_1a_2\left(k^2a_1a_2 - 2\right)}{1+k^2\left(a_1 + a_2\right)^2}\ \ .
\end{equation}
Since the transmission is supposed to be strong at SET, one must have $k^2a_1a_2 < 2$, according to Equation (11). Consequently, $T_a < T_b$, which implies that one must apply a higher voltage in order to SET the device in the case of (a). This has been confirmed experimentally, where the I---V curves under opposite polarities were found to be different. The average positive and negative $V_{SET}$ values are 2.164V and 1.652V, respectively \cite{26}.

In conclusion, the Mott transition in the anodic region of NiO RRAM has been described through the Hubbard model, given that the Hubbard $U$ is treated as an effective one. The screening effect renders a first order Mott transition in the solid as claimed by Mott, and a simple $ansatz$ has been introduced where the vacuum permittivity in the definition of $U$ is replaced by an effective permittivity. The nature of such transition is further shown in the Brinkman-Rice picture. The transition is established on a real electronic device structure, since the applied voltage triggers the local electron concentration variations inside NiO. In this way, we show that the transition should always be a local one. Accordingly, a possible asymmetric device behavior is predicted. The proposed mechanism has made the Mott transition practical in real electronic devices.  

\end{document}